\documentstyle[12pt,aaspp4,flushrt,twoside]{article}

\begin{document}

\markboth{D'Cruz, et al.}{HST Observations of HB structures in $\omega$ Cen}

\begin{flushright}
{\it Submitted to Astrophysical Journal}
\end{flushright}

\title{\bf HST Observations of New Horizontal Branch Structures in the
Globular Cluster $\omega$ Centauri\footnote{
Based on observations with the NASA/ESA Hubble Space Telescope,
obtained at the Space Telescope Science Institute, which is operated
by the Association of Universities for Research in Astronomy, Inc.
under NASA contract No.\  NAS5-26555. 
}}
\author{\sc Noella L. D'Cruz\altaffilmark{2}, Robert W. O'Connell, Robert T.
Rood, Jonathan H. Whitney, Ben Dorman\altaffilmark{3}}
\affil{Department of Astronomy, University of Virginia, P.O.Box 3818,
Charlottesville VA 22903; noella@physics.usyd.edu.au, rwo@virginia.edu,
rtr@virginia.edu, jw9w@virginia.edu, bd4r@virginia.edu}
\authoremail{noella@physics.usyd.edu.au}
\author{\sc Wayne B. Landsman, Robert S. Hill}
\affil{Laboratory for Astronomy \& Solar Physics, Code 681/ Raytheon
STX, NASA Goddard Space Flight Center, Greenbelt MD 20771; landsman@mpb.gsfc.nasa.gov,
bhill@uit.gsfc.nasa.gov}
\authoremail{}
\author{\sc Theodore P. Stecher}
\affil{Laboratory for Astronomy \& Solar Physics,  Code 681, NASA Goddard
Space Flight Center, Greenbelt, MD 20771; stecher@uit.gsfc.nasa.gov}
\authoremail{}
\author{\sc Ralph C. Bohlin}
\affil{Space Telescope Science Institute, Baltimore, MD 21218;
bohlin@fos.stsci.edu}
\authoremail{}
\altaffiltext{2}{Current Address: Chatterton Astronomy Department \& Research Centre for Theoretical Astrophysics, A-28 School of Physics, University of Sydney, NSW 2006, Australia}
\altaffiltext{3}{Current Address: Raytheon ITSS,
Code 644, NASA/GSFC, Greenbelt MD 20771; dorman@veris.gsfc.nasa.gov}


\begin{abstract}

The globular cluster $\omega$ Centauri contains the largest known
population of very hot horizontal branch (HB) stars.  We have used the
Hubble Space Telescope to obtain a far-UV/optical color-magnitude
diagram of three fields in $\omega$ Cen.  We find that over 30\% of
the HB objects are ``extreme'' HB or hot post-HB stars.  The hot HB
stars are not concentrated toward the cluster center, which argues
against a dynamical origin for them.  A wide gap in the color
distribution of the hot HB stars appears to correspond to gaps found
earlier in several other clusters.  This suggests a common mechanism,
probably related to giant branch mass loss.  The diagram contains a
significant population of hot sub-HB stars, which we interpret as the
``blue-hook'' objects predicted by D'Cruz et al.\ (1996a). These are
produced by late He-flashes in stars which have undergone unusually
large giant branch mass loss.  $\omega$~Cen has a well-known spread of
metal abundance, and our observations are consistent with a giant
branch mass loss efficiency which increases with metallicity.

\end{abstract}



\section{ \sc Introduction} 

Globular clusters contain the largest samples of old stellar
populations appropriate for studying late stages of stellar evolution,
especially the horizontal branch.  This consists of core
helium-burning, shell hydrogen-burning stars, with cores of
approximately the same mass but envelopes of a range of mass (Iben \&
Rood 1970).  For a given composition, envelope mass determines
horizontal branch (HB) temperature, with smaller envelopes yielding
higher temperatures.  The variation in envelope mass is due to
dispersion in mass loss during the preceding red giant branch (RGB)
phase.  However, the RGB mass loss process and the physical quantities
on which it depends are poorly understood.

Only with the advent of space-based imaging with the Hubble Space
Telescope (HST) and the {\it Astro}/Ultraviolet Imaging Telescope (UIT,
Stecher 1997) has it become feasible to study large samples of the
hottest (and lowest envelope mass) HB stars.  These objects are
brightest in the vacuum-UV band accessible to space telescopes.  But
equally important, globular cluster cores, which are impossible to
study from the ground because of image-crowding from cool
stars, can be studied from space because the cool stars are either
suppressed in the UV or resolved by the high spatial resolution in
the visible with HST.  Complete samples of hot stars can therefore be
obtained.

The hot HB populations of a number of globular clusters have recently
been studied, including $\omega$ Centauri, M3, M13, M79, M80, NGC 362,
NGC 2808, NGC 6388, NGC 6441, and NGC 6752 (e.g.\ Whitney et al.
1994; Moehler, Heber \& de Boer 1995; Hill et al.\ 1996; Landsman et
al.\ 1996; Ferraro et al.\ 1997; Sosin et al.\ 1997; Rich et al.\ 1997;
Dorman et al.\ 1997; O'Connell et al.\ 1997; Catelan et al.\ 1998;
Ferraro et al.\ 1998).  This work has revealed several important
features of the hot horizontal branch.  {\it (i)} Some HBs extend
nearly all the way to the He-burning main sequence, having
temperatures $\ga 16000\,$K and envelope masses of $\lesssim 0.05
M_\odot$, presumably due to extreme RGB mass loss rates.  This region
is called the ``extreme horizontal branch'' (EHB).  {\it(ii)} Some
clusters have significant populations of hot stars lying above the
HB. These are probably post-EHB objects which, because of their small
envelopes never reach the thermally pulsing stage on the asymptotic
giant branch (AGB) and instead spend most of their post-HB lifetime at
high temperatures. EHB descendents evolve along either post-early AGB
tracks or ``AGB-manqu\'{e}'' tracks (e.g.\ Greggio \& Renzini 1990). {\it
(iii)} Some HBs have significantly underpopulated regions or ``gaps,''
and these can occur at similar temperatures in different clusters
(Ferraro et al.~1998).  {\it (iv)} In the case of $\omega$ Cen, there
appear to be hot stars lying below the HB (Whitney et al.\ 1994).  All
of these features are probably related to physical processes on the
RGB that determine the distribution of HB envelope masses.

In this paper, we discuss new far-UV/visible HST observations of the
globular cluster $\omega$ Centauri, which is an important benchmark
system in several ways.  It exhibits a spread in metallicity: $-2.2
\lesssim {\rm [Fe/H]} \lesssim -0.8$ (Norris, Freeman, Mighell 1996,
Greenlaw 1993, Woolley 1966), which makes it a better analogue to the
old populations of external galaxies than most local clusters.
(Evidence for multiple generations of star formation suggest that
$\omega$ Cen itself may even be a captured galaxy nucleus, Hughes \&
Wallerstein 1999.)  Its HB is extremely blue, and it contains the
largest identified population of EHB and of putative EHB-descendent
stars lying above the HB (Dickens et al.~1988, Bailyn et al.~1992,
Whitney et al.~1994).  We are particularly interested in investigating
the reality of two unusual HB features found in lower-precision far-UV
photometry by the UIT: a hot HB gap and sub-HB stars (Whitney et al.\
1994).

\section{\sc Observations and Data Reduction}

We have used HST to observe $\omega$ Cen in the far-UV and visual
bands as part of GO program 6053.  The images discussed here are Wide
Field Planetary Camera 2 (WFPC2) ``acquisition frames'' taken to
confirm astrometry for a set of spectroscopic targets (to be discussed
elsewhere) identified on UIT far-UV images by Whitney et al.\ (1994).
Compared to the Whitney et al.\ study, the HST data has smaller
photometric errors, better astrometry, and an optical comparison image
better matched to the UV.  Three fields in $\omega$ Cen were observed
with WFPC2 in the F160W and F555W filters.  Total exposure times for
each field were 1400 secs in F160W and 8 sec in F555W,  each split into
two parts to allow the elimination of most cosmic ray events via
a simple combination program.  The fields
were located 1\farcm 34, 2\farcm 62, and 4\farcm 57 from the
cluster center (${\rm R.A. = 13^h\, 26^m\, 45\fs 9}$; ${\rm DEC} =
-47\arcdeg \,28 \arcmin \, 37 \arcsec$, J2000). The choice of the
three fields is based on the position of stars for which UV spectra
were obtained.

\begin{figure*}[bhtp]
\vskip3.0truein
\includegraphics{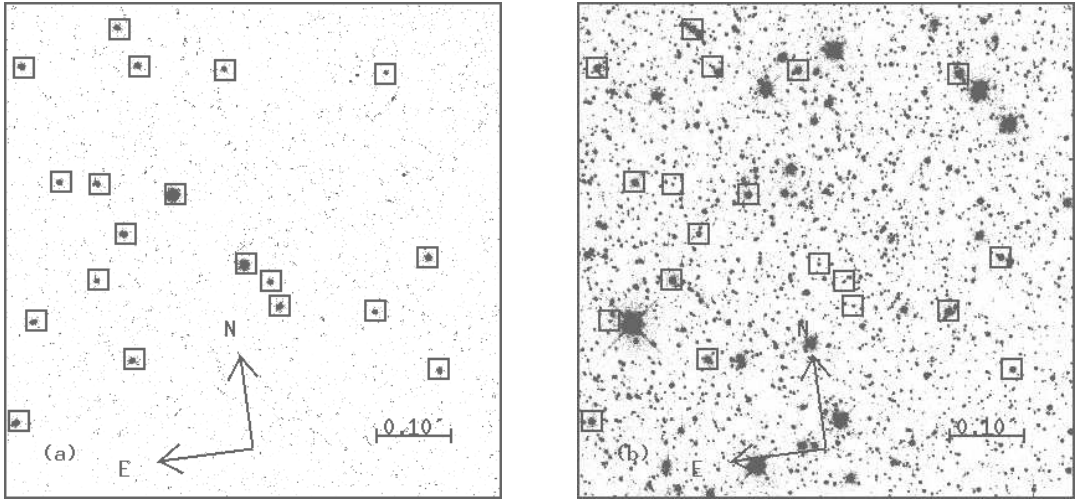}
\figcaption[fig1]{\small (a) Far-UV and (b) $V$ frames of part of our
HST/WFPC2 field at a radius of 2\farcm 62 from the center of $\omega$
Cen.  UV sources and their visual counterparts are marked. The boxes
are centered on the $V$ band centroids, and the offsets between the
stars and the $V$-centroids in (a) shows the extent of geometric
distortion.
\label{fig:fig1}}
\end{figure*}

Figure~1 compares the far-UV and $V$ images of part of one of our
fields.  There are about 100 times as many sources present on the
F555W frame as on the F160W frame.  F555W counterparts could be
identified for most of the F160W sources in the field if the geometric
distortion in the far-UV filter is taken into account (Biretta 1996).
The identifications are marked on both frames in Fig.\ 1 with the
boxes centered on the $V$ band centroids.  The offsets between the box
centers and the star images on the far-UV frame illustrate the extent
of the image distortion.  A brief inspection of the UV frames
indicates that there is vignetting in the filter and that the images
are affected by coma, but there are no blended images.  Aperture
photometry was carried out on the UV sources using an IDL
implementation of DAOPHOT (Stetson 1987).  The area around each star
was cleaned to make sure that no cosmic rays or nearby stars
contaminated the sky magnitude.  Contamination effects on our UV
images (Holtzman et al.~1995) were minimal since our observations were
taken only one day after a decontamination cycle.  

Obtaining stellar magnitudes from the $V$ band frames required point
spread function (PSF) fitting, for which we used the FORTRAN version
of DAOPHOT II (Stetson 1992). Aperture corrections were assumed to be
constant with magnitude for the recommended aperture of 1\arcsec\ in
diameter (Holtzman et al.~1995). Results were converted into the STMAG
magnitude system as described by Holtzman et al. For a monochromatic
flux, $F_\lambda$ (in units of $ {\rm erg\, s}^{-1}\,{\rm
cm}^{-2}\,$\AA$^{-1}$) , at a wavelength $\lambda$, ${\rm STMAG} =
-2.5 \times \log F_\lambda - 21.1$.

\vspace*{3mm}

\section{\sc Results and Discussion}

\subsection{\em The UV-Optical CMD}

Figure~2 shows the resulting $m(160),(160-V)$ color-magnitude diagram for
the three fields combined. There are 406 stars in the CMD from a total
of 473 UV sources in the three fields. Of the 67 sources that do not
appear in the CMD, 3 are on bad columns in the UV, 21 are either near
the edge of the chip or too close to another star to measure in the
UV, 8 have no $V$ counterpart (these may not be stellar), and 35
(mostly in the more crowded central region) are too faint to measure
accurately in $V$. Many of these 35 could be EHB and AGB-manqu\'{e}
stars since they are relatively bright in the UV. The detection limit
is the dot-dashed line on the figure.

\begin{figure*}[htbp]
\vskip4.5truein
\includegraphics{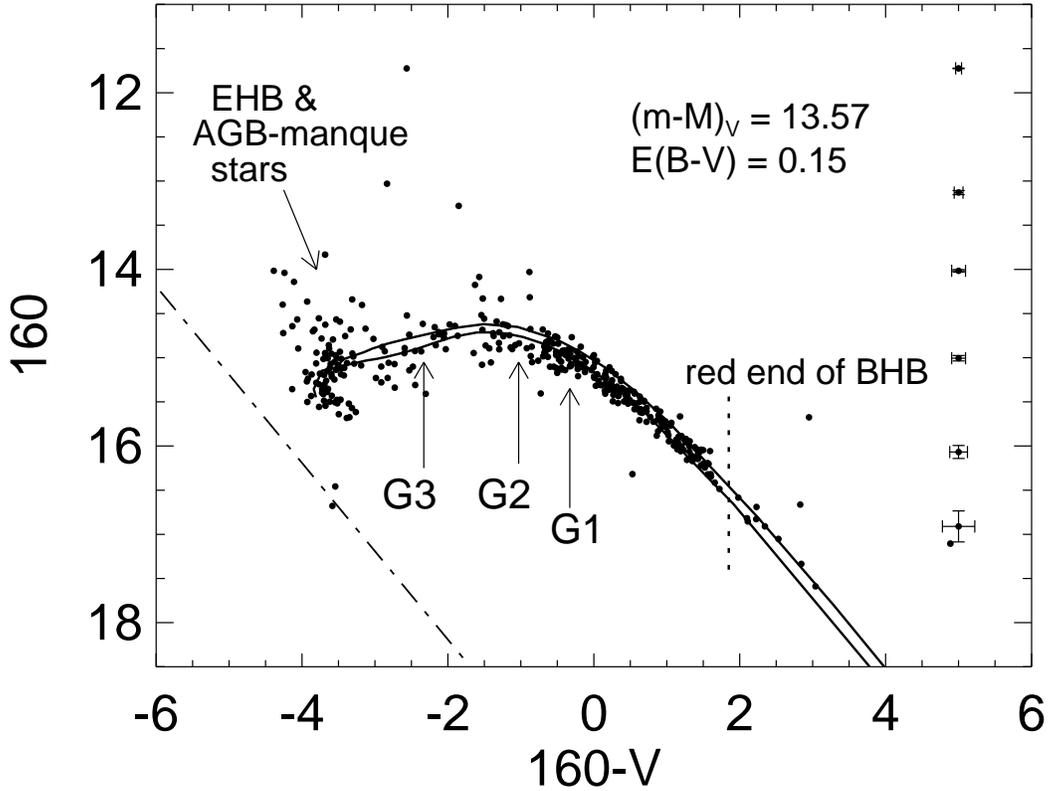}
\figcaption[cmds] {\small The observed $m(160),(160-V)$ color-magnitude diagram
for 406 stars in $\omega$ Centauri. The upper solid line is a ZAHB for
${\rm [Fe/H]}= -2.2$, while the lower line is for ${\rm [Fe/H]}= -1.5$
from Dorman (1997), transformed using the distance modulus and
extinction given in the figure. The upper detection limit is shown by
the dot-dashed line. Typical photometric error bars are shown at the
right.  There is a large sub-HB population most conspicuous at the hot
end of the HB but extending several magnitudes in $(160-V)$. A number of
AGB-manqu\'{e} or other post-HB stars are seen above the HB. The
approximate position of the red end of the BHB is marked. The
locations of the HB gaps, G1, G2, G3, found by Ferraro et al.~(1998)
in M13 and M80, are shown.
\label{fig:cmds}}
\end{figure*}

\begin{figure*}[bhtp]
\vskip7.0truein
\includegraphics{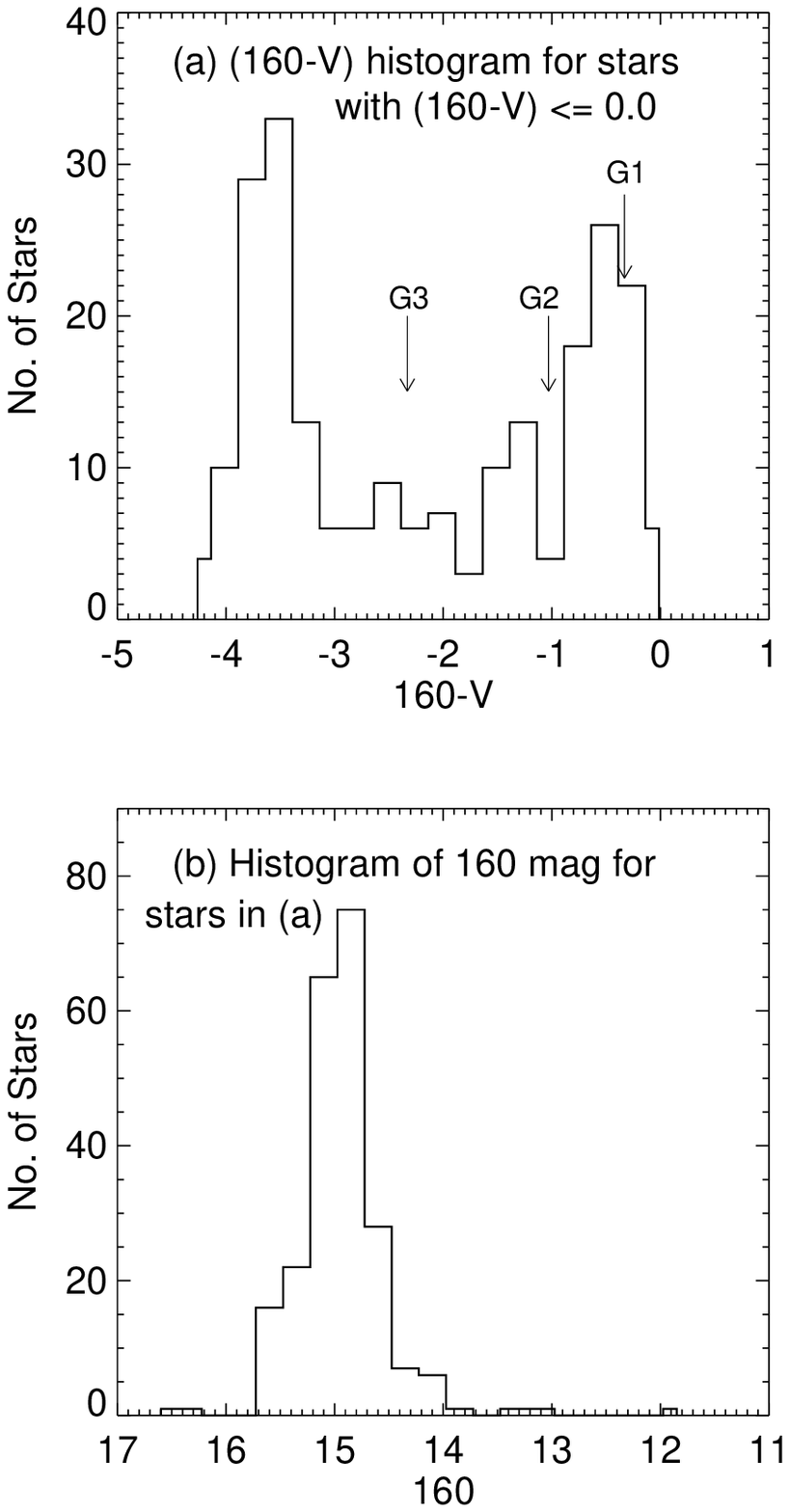}
\figcaption[fig3]{\small (a) Histogram of $(160 - V)$ colors for stars on the
``horizontal'' part of the hot HB in $\omega$ Cen bluer than $(160 -V)
= 0.0$.  The locations of HB gaps, G1, G2, G3, found by Ferraro et
al.~(1998) in M13 and M80, are shown.  (b) Histogram of $m(160)$
magnitudes for the stars in (a), showing the presence of both post-HB
evolved stars (brighter than the HB) and sub-HB stars. }
\end{figure*}

Two zero-age HBs (ZAHBs) have been plotted on the observed color
magnitude diagram, corrected for $\omega$ Cen's distance modulus of
13.57 (Dickens, et al.~1988) and interstellar reddening of $E(B-V) =
0.15$ taken from Whitney et al.~(1994) and Greenlaw (1993). These
values produce the best overall match between the theoretical and
observed sequences. We use the UV extinction law of Cardelli, Clayton
\& Mathis (1989). The upper solid line is the ZAHB for ${\rm [Fe/H]} =
-2.2$ while the lower line is for ${\rm [Fe/H]} = -1.5$.  Typical
errors in the data are shown on the right side of Figure 2.

The main features of interest in the CMD in Fig.\ 2 are the following:

\begin{enumerate}

\item The cooler HB stars with $-0.5 \lesssim (160 - V) \lesssim 2.0$
($T_{\rm eff} \sim$ 8000 to $11000\,{\rm K}$) fit the predicted locus
of the ZAHB well, with only a small difference in mean slope (the data
being slightly flatter than the models). They show relatively little
vertical scatter. By contrast, the hotter objects at comparable $m(160)$
magnitudes show large scatter, which must therefore be cosmic, not
photometric errors.

\item The approximate position of the cool end of the heavily
populated segment of the blue HB is marked on the CMD. The limiting
magnitude of the UV images is significantly fainter than this, so the
reduced population is not a detection artifact.  The boundary lies at
$T_{\rm eff} \sim 7600\,$K, based on the Dorman models.  Stars redder
than this tend to lie above the HB and are probably evolved BHB stars
rather than ZAHB stars.

\item About 32\% of the HB sample is composed of EHB stars,
with $(160-V) \lesssim -1.9$.  
The objects lying more than 0.2 mag above the predicted EHB are
probably AGB-manqu\'{e} descendents of EHB stars.  A total of 389 such
objects were found in the larger field sampled by UIT.

\item There is a large sparsely populated region in the HB from $-3.0
\lesssim (160-V) \lesssim -0.5$ ($22500 \, {\rm K}\lesssim T_{\rm eff}
\lesssim 11000 \, {\rm K}$).  This is shown more clearly in Figure~3(a)
which plots the $(160-V)$ distribution for stars with $(160-V) \le 0$
(where the HB is approximately horizontal).  This underpopulated
region corresponds to the part of the HB where Ferraro et al.~(1998)
found gaps in M13, M80 and M3. We have marked the approximate position
of the gaps G1, G2, and G3 from Figure~4 of Ferraro et al.~in both
Figs.\ 2 and 3(a) after correcting for the reddening difference between
the clusters.  It appears that there is a broad gap in $\omega$ Cen
corresponding to G3 and a narrower one corresponding to G2 (though the
statistics are not good) but that G1 is clearly absent. 

\item The HST observations confirm, with good photometric precision,
the presence of sub-HB stars, as found by UIT.  A histogram of $m(160)$
magnitudes for the hot HB is shown in Figure 3(b).  The mode of the
blue HB occurs at $m(160) = 15.2$, and the expected post-HB objects
lie at brighter magnitudes.  However, there is a tail of fainter
objects up to 0.7 mags fainter.  Most are concentrated near the hot
end of the EHB; however, other sub-HB objects are found at colors as
red as $(160 -V) \sim -0.5$. 

\end{enumerate}

\subsection{\em Spatial Distribution of Hot Stars}

There is no evidence from their spatial distribution that the EHB
objects in $\omega$ Cen are binary stars or that dynamical effects are
important in their production.  Binaries or the products of strong
interactions in the cluster core would tend to be more concentrated to
the cluster center.  Instead, in the large-field (40\arcmin\ diameter)
UIT images of (Whitney et al.~1994), the radial distribution of
$\omega$ Cen's EHB stars follows that of the rest of the HB stars.
Our HST data suggest that the fraction of EHB stars increases outwards
from about $28-33\%$ in the inner two fields to about 43\% in the
outer field.  This relatively modest effect is possibly an artifact of
our inability to obtain $V$-band magnitudes for the fainter hot
objects in the cluster's interior.

\subsection{\em Horizontal Branch Gaps}

It is likely that the HB gaps apparent in Figs.\ 2 and 3(a) are
connected to the distribution of envelope masses produced by the RGB
mass loss process.  Rood, Whitney \& D'Cruz~(1997) have shown that HB
gaps cannot arise from post-ZAHB evolution if the distribution of
envelope masses is uniform. 

Whitney et al.~(1998) modeled the broad G3 gap which was detectable in
the earlier UIT data using smooth distributions in total ZAHB mass or
in the mass loss efficiency parameter $\eta$ (from Reimers' mass loss
formula, Reimers 1977).  They incorporated the models of D'Cruz et
al.\ (1996a), in which the effects of mass loss (as specified by
$\eta$) on RGB evolution are followed in detail and the ``blue hook''
population is included.  They found that a flat distribution in
$\eta$, when combined with the known distribution of metallicity, best
reproduced the broad G3 gap and the sub-HB stars in the UIT data.
The metallicity spread would presumably act to blur finer structure
in the $\omega$ Cen HB, whatever the mechanism for its origin.  

Ferraro et al.~(1998) suggest that gaps are located at roughly the
same effective temperatures in different clusters, though not all
clusters have the same number of gaps. The hottest gap, G3, is found
in extreme ``blue tail'' clusters (Fusi Pecci et al.~1993), and may
indicate the onset of EHB behavior.  The $\omega$ Cen HB
extends further to the blue than the HBs of M13 and M80. This could
imply that mass loss efficiency is greater in $\omega$ Cen than in M13
and M80.  The Whitney et al.\ model does not explain the presence of
multiple gaps, which suggest the existence of several mass loss
mechanisms operating with different efficiencies in different
clusters.  Sweigart's (1997) models which involve rotationally-induced
envelope mixing of He on the RGB, predict larger amounts of mass loss
due to increased luminosity of the red giant tip.  It is possible that
the onset of increased mixing could manifest itself on the HB as a gap.

\subsection{\em Sub-HB Stars and Metallicity Dependence of Mass Loss}

$\omega$ Cen is the only cluster in which there is now clear evidence
for sub-HB objects.  Whitney et al.\ (1998) interpreted the sub-HB
objects in the UIT sample as the ``blue hook'' stars predicted in the
models of D'Cruz et al.\ (1996a). These are objects which, due to
extreme amounts of RGB mass loss, left the RGB before the He flash but
were able to ignite He later as they started moving down the high
temperature part of the white dwarf sequence. They are predicted to lie
at the very blue end of the ZAHB.  Their core masses could be as much
as $0.015 \, M_\odot$ below canonical values, which will cause them to
lie up to $0.1 \, {\rm mag}$ below the ZAHB.  The computations of
Sweigart (1999) suggest that the blue hook stars might have higher
temperatures ($\log T_{\rm eff} \sim 4.6$) than typical EHB stars, and
surface abundances enriched in helium and carbon.  At these high
temperatures, even the $(160 - V)$ colors do not provide good
$T_{\rm eff}$ discrimination, so that followup spectroscopy would be
useful to provide constraints on both the temperatures and the
abundances.

Some of the sub-HB stars are also likely to be affected by radiative
levitation.  The lowest temperature at which they appear in the CMD is
$T_e \sim 11,000$ K, roughly
the temperature expected for the onset of radiative levitation of iron,
which would cause them to be up to another $0.1\,{\rm mag}$ fainter in
$m(160)$ (Grundahl et al.\ 1999, Behr et al.\ 1999, Moehler et
al.\ 1999).  

However, the stars in Fig.~2 lie farther below the ZAHB
than predicted by the models. 
The metallicity spread in $\omega$ Cen may account for some of the
differences between the observed and predicted CMDs. Metal poor HB
stars are more luminous than metal rich HB stars of the same $T_{\rm
eff}$. The observed CMD would be more consistent with the models if
the metallicity of the HB stars {\em increases} with increasing
effective temperature. This would produce the flatter slope for the
observed CMD than for a single metallicity theoretical ZAHB,  just as
observed for $(160-V) \ga -1$. In this interpretation, the sub-HB stars
would mostly be metal rich blue hook stars.

If, indeed, metallicity increases with effective temperature along the
cluster's HB, then this suggests that mass loss efficiency increases
with metallicity. This result is consistent with the determination of
[Ca/H] in the cluster's cooler BHB stars, where a large fraction of
the cooler BHB stars with $ 7500\, {\rm K} \lesssim T_{\rm eff}
\lesssim 8200 \, {\rm K}$ are found to be metal poor
($\langle {\rm [Fe/H]} \rangle = -1.95$) (D'Cruz et al.\ 1996b;
D'Cruz 1998). The calcium abundance distribution of the giants and RR
Lyraes is peaked at the higher metallicity of ${\rm [Fe/H]} = -1.7$,
possibly indicating that HB stars of this metallicity or higher
dominate the abundance distribution at higher temperatures. 

Finally, we note that a mass loss efficiency which increases with
metallicity, as suggested by our data, is in the correct sense to
explain the well-known correlation between far-UV output and metal
abundance in elliptical galaxies (Burstein et al. 1988, Greggio \&
Renzini 1990; Dorman, O'Connell, \& Rood 1995; Yi, Demarque,
\& Oemler 1997).  

\section{\sc Conclusion}

HST observations of $\omega$ Cen confirm the substantial population of
sub-HB, EHB, and AGB-manqu\'{e} stars found by Whitney et al.~(1994)
with UIT. About 32\% of the HB consists of EHB stars and their
descendents.  The density of EHB objects relative to the total HB does
not decrease with radius, contrary to expectations if they are in
binary systems or if dynamical interactions in the cluster core are
critical for their production.  The sub-HB stars are probably
``blue-hook'' stars produced by RGB stars with large mass loss which
experience He ignition after leaving the RGB.  A broad gap in the blue HB
seen in earlier work is confirmed and corresponds to two of the hotter
gaps found in the clusters M13 and M80, possibly blurred by $\omega$
Cen's metallicity spread.  The gaps probably contain information on
RGB mass loss mechanisms.  

Comparison of the observed CMD to theoretical ZAHB models suggests
that RGB mass loss efficiency may increase with metallicity and that
the sub-HB stars are the most metal-rich. This is in the correct sense
to explain the well-known correlation between far-UV output and metal
abundance in elliptical galaxies.

\noindent {\bf Acknowledgements:} We are grateful to Barbara
Paltrinieri, Francesco Ferraro, and Jamie Ostheimer for their help with
the data reduction.  This research was supported in part by STScI
grant GO-06053 and NASA grants NAG5-700 and NAG5-6403.


\end{document}